\documentclass[%
reprint,
 superscriptaddress,preprintnumbers,
 nofootinbib,
 amssymb,
 aps,
]{revtex4-2}

\usepackage[a4paper, margin=0.5in]{geometry}

\usepackage[utf8]{inputenc}
\usepackage{hyperref}
\usepackage{amsmath,amssymb,mathtools}
\usepackage{dsfont}
\usepackage{graphicx}   
\usepackage[table]{xcolor}
\usepackage{colortbl}
\usepackage{url}
\usepackage[normalem]{ulem}
\usepackage{slashed}
\usepackage[capitalise, english]{cleveref}
\usepackage[noindentafter]{titlesec} 
\allowdisplaybreaks
\usepackage{comment}
\usepackage{bm}
\usepackage{booktabs}
\usepackage{physics}
\usepackage{ragged2e}
\usepackage{multirow}
\usepackage{makecell}

\usepackage{lipsum}

\usepackage{siunitx}
\usepackage{xspace}
\usepackage{enumitem}
\usepackage{subcaption}
\usepackage{tikz}
\usetikzlibrary{calc,tikzmark,fit,shapes.geometric,matrix,decorations.markings,arrows.meta,decorations.pathmorphing,patterns,positioning,snakes}

\usepackage[font=small,labelfont=bf]{caption}
\tikzset
  {midarrow/.style={decoration={markings,mark=at position 0.5 with
     {\arrow[thin,xshift=2pt]{Triangle[length=4pt,#1]}}},postaction={decorate}}
  }

\tikzset{
proton/.style = {circle, draw=black, thin, fill=black!20!white, minimum size=#1,
              inner sep=0pt, outer sep=0pt},
proton/.default = 6pt 
}

\tikzset{
blob/.style = {circle, draw=black, thin, preaction={fill, black!20!white}, pattern=north east lines, minimum size=#1,
              inner sep=0pt, outer sep=0pt},
blob/.default = 6pt 
}

\tikzset{
wc/.style = {circle, fill, minimum size=#1,
              inner sep=0pt, outer sep=0pt},
wc/.default = 4pt 
}

    \makeatletter
    \def\CT@@do@color{%
      \global\let\CT@do@color\relax
            \@tempdima\wd\z@
            \advance\@tempdima\@tempdimb
            \advance\@tempdima\@tempdimc
    \advance\@tempdimb\tabcolsep
    \advance\@tempdimc\tabcolsep
    \advance\@tempdima2\tabcolsep
            \kern-\@tempdimb
            \leaders\vrule
                    \hskip\@tempdima\@plus  1fill
            \kern-\@tempdimc
            \hskip-\wd\z@ \@plus -1fill }
    \makeatother

\usepackage{accents}
\DeclareMathSymbol{\widetildesym}{\mathord}{largesymbols}{"65}

\titleformat*{\subsubsection}{\normalfont \small \bfseries \boldmath}
\renewcommand{\paragraph}[1]{\vspace{.3em} \indent {\bfseries \boldmath #1 ---}\xspace }
\makeatletter
    \renewcommand{\p@subsection}{}
    \renewcommand{\p@subsubsection}{}
\makeatother

\colorlet{mylinkcolor}{red}
\colorlet{mycitecolor}{blue}
\colorlet{myurlcolor}{red}

\hypersetup{
  linkcolor  = mylinkcolor,
  citecolor  = mycitecolor,
  urlcolor   = myurlcolor,
  colorlinks = true
}
\DeclareUnicodeCharacter{2212}{\ensuremath{-}}


\begin{document}

\title{
 \boldmath Cornering Natural SUSY at a Tera-$Z$ Factory
}
\author{Admir Greljo}
\email{admir.greljo@unibas.ch}
\affiliation{Department of Physics, University of Basel, Klingelbergstrasse 82, CH-4056 Basel, 
Switzerland}
\author{Ben A. Stefanek}
\email{bstefan@ific.uv.es}
\affiliation{Instituto de F\'isica Corpuscular (IFIC), Consejo Superior de Investigaciones \\Cient\'ificas (CSIC) and Universitat de Val\`{e}ncia (UV), 46980 Valencia, Spain}
\author{Alessandro Valenti}
\email{alessandro.valenti@unibas.ch}
\affiliation{Department of Physics, University of Basel, Klingelbergstrasse 82, CH-4056 Basel, 
Switzerland}



\begin{abstract}

The future circular $e^+ e^-$ collider (FCC-ee) stands out as the next flagship project in particle physics, dedicated to uncovering the microscopic origin of the Higgs boson. In this context, we assess indirect probes of the Minimal Supersymmetric Standard Model (MSSM), a well-established benchmark hypothesis, exploring the complementarity between Higgs measurements and electroweak precision tests at the $Z$-pole. We study three key sectors: the heavy Higgs doublet, scalar top partners, and light gauginos and higgsinos, focusing on the parameter space favored by naturalness. Remarkably, the Tera-$Z$ program consistently offers significantly greater indirect sensitivity than the Mega-$h$ run. While promising, these prospects hinge on reducing SM uncertainties. Accordingly, we highlight key precision observables for targeted theoretical work.

\end{abstract}

{
\hypersetup{linkcolor=black, urlcolor=black}
\maketitle
}

\section{Introduction} 
\label{sec:intro}

Understanding the microscopic origin of the Higgs boson remains a central motivation for exploring physics at the TeV scale and beyond, shaping the scientific goals of present and future colliders. In this context, the Minimal Supersymmetric Standard Model (MSSM) remains one of the best motivated theoretical frameworks~\cite{Haber:1984rc, Martin:1997ns, Dreiner:2023yus, Allanach:2024suz}, serving as a key benchmark for testing the principle of naturalness.

The future direction for particle physics beyond the HL-LHC era is actively debated, notably within the ongoing European Strategy for Particle Physics (ESPP)~\cite{EuropeanStrategyforParticlePhysicsPreparatoryGroup:2019qin, Altmann:2025feg}. Several proposed flagship collider facilities, including circular colliders FCC-ee~\cite{FCC:2025lpp}, CEPC~\cite{CEPCStudyGroup:2018ghi}, LEP-3~\cite{Anastopoulos:2025jyh}, and linear colliders ILC~\cite{ILC:2013jhg}, CLIC~\cite{CLICdp:2018cto}, LCF~\cite{LinearCollider:2025lya}, are primarily motivated by precision Higgs boson studies (\textit{Higgs factories}). Circular colliders uniquely offer a complementary run at the $Z$-pole, where the high luminosity allows for the production of over 1 trillion $Z$ bosons. The extraordinary statistical power of the Tera-$Z$ run provides complementary or better sensitivity compared to the Mega-$h$ run, enabling precise probes of subtle quantum effects of new physics~\cite{Allwicher:2023shc, Allwicher:2024sso, Greljo:2024ytg, Maura:2024zxz,  Stefanek:2024kds, Erdelyi:2024sls, Gargalionis:2024jaw, Celada:2024mcf, Davighi:2024syj, Erdelyi:2025axy, Maura:2025rcv, Bordone:2025cde, terHoeve:2025gey, Allwicher:2025bub, deBlas:2021jlt, deBlas:2022ofj,Azzi:2025dwl}.

In this letter, we study the MSSM as a well-established benchmark for a microscopic theory of the Higgs boson to show that electroweak precision observables at FCC-ee offer complementary (and often superior) sensitivity to natural supersymmetry (SUSY) compared with Higgs coupling measurements. While previous studies have explored related directions~\cite{Fan:2014vta, Fan:2014axa, Knapen:2024bxw, Nagata:2025ycf}, our work introduces three key novelties.

First, our analysis leverages the recently published Feasibility Study Report (FSR) for FCC-ee~\cite{FCC:2025lpp}, which outlines a detailed experimental program and provides state-of-the-art projections for a wide range of electroweak observables (see Table~2 in~\cite{FCC:2025lpp}). These updated inputs allow a more realistic evaluation of the sensitivity.

Second, the projected experimental precision at FCC-ee sets stringent requirements on the accuracy of Standard Model (SM) theory predictions, with many observables becoming limited by theoretical uncertainties. Since future theoretical progress is difficult to anticipate, we assess the impact of theory uncertainties on electroweak precision observables by considering three scenarios: S1 (conservative theory progress), S2 (aggressive theory progress), and S3 (experimental errors only). For the less precise Higgs observables, we assume theoretical errors will be negligible. These scenarios are motivated by ongoing community discussions and are presented in the Appendix. Comparing theory limited scenarios to the case of experimental errors only allows us to identify key observables where improved SM theory calculations would yield the most benefit, highlighting priorities for future theoretical efforts.

Finally, we focus on three key sectors that provide complementary probes of natural supersymmetry at FCC-ee: additional Higgs bosons (Section~\ref{sec:heavyHiggs}), scalar top partners (Section~\ref{sec:stops}), and gauginos/higgsinos (Section~\ref{sec:gaugino}). To enable a light superpartner spectrum consistent with naturalness, we assume a suppression of flavor and CP violation through a mechanism such as Minimal Flavor Violation (MFV)~\cite{DAmbrosio:2002vsn}. This assumption renders current bounds from flavour observables, including $b \to s\gamma$~\cite{Misiak:2015xwa}, subdominant compared to the robust constraints we will derive. A detailed assessment of flavour measurement projections is left to future work. The degree of electroweak fine-tuning is quantified using the Barbieri–Giudice measure~\cite{Barbieri:1987fn}, defined as $\Delta_a = \partial \log m_Z^2/\partial \log a$, where $a$ denotes a parameter of the MSSM Lagrangian.

\section{Heavy Higgs doublet} 
\label{sec:heavyHiggs}

\begin{figure}[h!]
    \centering
        \begin{tikzpicture}[thick,>=stealth,scale=1.2,baseline=-0.5ex]
            \draw[midarrow] (-1.9,0.8) node[left] {$b_R$} -- (-0.8,0.8);
            \draw[midarrow] (-0.8,0.8) -- (0.8,0.8);
            \draw[midarrow] (0.8,0.8) -- (0.8,-0.8);
            \draw[midarrow] (0.8,-0.8) -- (-0.8,-0.8);
            \draw[midarrow] (-0.8,-0.8) -- (-1.9,-0.8) node[left] {$\bar b_R$};
            \draw[dashed] (-0.8,0.8) -- (-0.8,-0.8);
            \draw[dashed] (0.8,0.8) -- (1.8,0.8) node[right] {$H^\dagger$};
            \draw[dashed] (0.8,-0.8) -- (1.8,-0.8) node[right] {$H$};
            \node[above] at (0,0.8) {$q_L$};
            \node[below] at (0,-0.8) {$q_L$};
            \node[right] at (0.8,0) {$t_R$};
            \node[left] at (-0.8,0) {$\Phi$};
        \end{tikzpicture}
\caption{Log-enhanced $Zb_R \bar b_R$ correction from the heavy MSSM Higgs doublet $\Phi$, generating $[Q_{Hd}]_{33}$ in the decoupling limit. See Section~\ref{sec:heavyHiggs} for details}
    \label{fig:ZbbBox}
\end{figure}
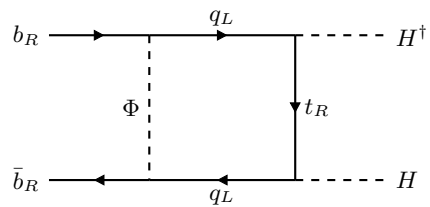

\begin{figure*}[!htb]
    \centering
    \includegraphics[scale=0.985]{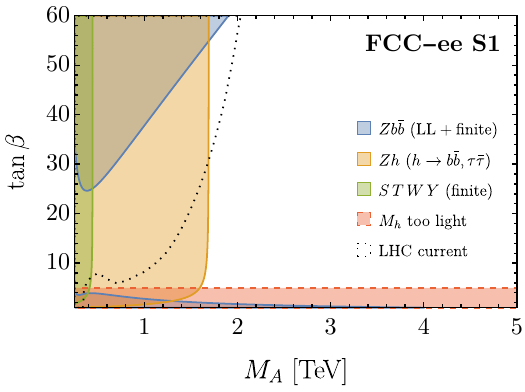} \hspace{3mm}
    \includegraphics[scale=1]{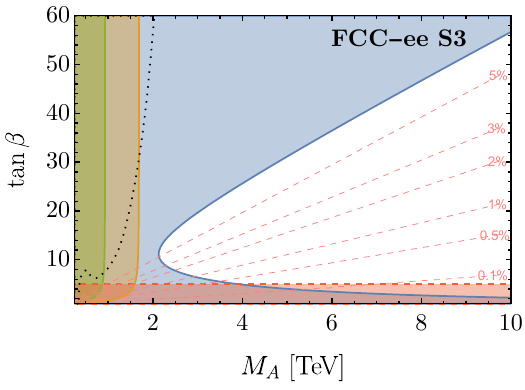}
    \caption{\justifying FCC-ee projected $2\sigma$ sensitivity to the heavy MSSM Higgs doublet in the decoupling limit, as functions of $M_A$ and $\tan\beta$. The left panel shows the conservative S1 theory scenario, while the right panel assumes the aggressive S3 scenario. Shaded regions correspond to expected sensitivities from $Z\to b \bar b$ (blue), Higgs decays (orange), and oblique electroweak parameters (green). The black dotted line marks current LHC direct search limits~\cite{ATLAS:2020zms}. Dashed red lines in the right panel indicate fine-tuning contours. The red-shaded region requires heavy stops to fit the Higgs mass, adding additional fine-tuning. See Section~\ref{sec:heavyHiggs} for details.}
    \label{fig:Higgs}
\end{figure*}

The heavy Higgs doublet $\Phi$ occupies a unique role in the $R$-parity-conserving MSSM: it is the only new field that couples linearly to the SM.  In the decoupling limit ($\beta - \alpha = \pi/2$)~\cite{Dreiner:2023yus}, tree-level matching onto the dimension-6 SM Effective Field Theory (SMEFT) includes the operators $Q_{uH}$, $Q_{dH}$, and $Q_{eH}$~\cite{deBlas:2017xtg}.  Their Wilson coefficients (WCs) depend solely on the SM Yukawa matrices, $\tan\beta$, and the quartic coupling of the scalar potential term $(\Phi^\dagger H + H^\dagger \Phi)\,|H|^2$, which in the MSSM is entirely fixed by the gauge couplings and $\tan\beta$.  These operators shift Higgs-fermion interactions, inducing deviations in Higgs decays. Future Higgs factories, with their sub-percent precision on Higgs couplings, will therefore place stringent, model-independent bounds on the heavy Higgs doublet.

However, one-loop effects at Tera-$Z$ can also yield competitive constraints on new physics typically probed by above-pole observables, thanks to the enormous $Z$-statistics available \cite{Maura:2024zxz, Greljo:2024ytg}.  These loop corrections split into two classes: $i)$ flavor universal (oblique) modifications of the SM gauge boson propagators, traditionally encoded in the $S, T, W, Y$ parameters~\cite{Peskin:1990zt, Barbieri:2004qk}, and $ii)$ non-oblique, Yukawa-dependent corrections to the $Z$-fermion vertices.  Of particular importance are shifts in the $Zb_R \bar b_R$ coupling, described in the SMEFT by the operator $Q_{Hd}$, generated by $\Phi$ at one loop via the box graph in~\cref{fig:ZbbBox}. Defining $M_A$ to be the mass of $\Phi$, the WC of this operator at the electroweak scale includes a $\tan \beta$ and log-enhanced contribution
\begin{align}
[C_{Hd}]_{33}(M_Z) \supset -\frac{y_t^2 y_b^2 \tan^2\beta}{32 \pi ^2 M_{A}^2}\left(1+\log \frac{M_Z^2}{M_A^2}\right) \,,
\end{align}
which renders this effect very important at large $\tan \beta$.
Modifications to left-handed bottom couplings are encoded in $Q^{(1,3)}_{Hq}$, but their impact is significant only for small $\tan\beta$, a region less relevant for natural SUSY since obtaining the observed Higgs mass would require very heavy stops.  Indeed, the shift of the $Zb_L\bar b_L$ vertex proportional to
$[C_{Hq}^{(1)}+C_{Hq}^{(3)}]_{33} \supset 
y_t^4 \cot^2 \beta/(32\pi^2 M_A^2) \left(1+\log M_Z^2 / M_A^2\right)$, becomes negligible for large $\tan\beta$ despite the $y_t^4$ dependence.
The logarithmic part of these non-oblique corrections is scheme independent---it can be understood as the 4-quark operators $C_{qu}^{(1)} \propto y_t^2 \cot^2\beta$ and $C_{qd}^{(1)} \propto y_b^2 \tan^2\beta$ generated by integrating out $\Phi$ at tree level running with $y_t^2$ into $C_{Hq}^{(1)}$ and $C_{Hd}$, respectively~\cite{deBlas:2017xtg,Jenkins:2013wua}. As we will see, the unprecedented precision on $R_b$ at Tera-$Z$ ensures that $Q_{Hd}$ provides the leading bound in our most aggressive S3 benchmark (which assumes no theoretical limitations).

To quantitatively assess these effects, we perform the full one-loop matching onto the dimension-6 SMEFT by integrating out the heavy Higgs doublet in the decoupling limit.  The matching is implemented with a custom MSSM model in \texttt{Matchete} \cite{Fuentes-Martin:2022jrf} that we have cross-checked against the results of \cite{Kraml:2025fpv}.  We then construct a global likelihood that incorporates the contributions of all relevant dimension-6 operators to precision observables at FCC-ee, including electroweak precision tests at the $Z$- and $W$-poles (see \cref{app:EWPO}), Higgs coupling measurements \cite{FCC:2025lpp}, and fermion pair-production ratios above the $Z$-pole \cite{Greljo:2024ytg}. Observables are computed at tree level in the SMEFT except at the $Z$- and $W$-pole, where we include 1-loop corrections~\cite{Bellafronte:2023amz, Biekotter:2025nln}. Since this NLO computation is given at the scale $M_Z$, we make the same choice in our one-loop matching calculation which corresponds to performing the leading-log renormalization group evolution (RGE) from the scale $M_A$ to $M_Z$, canceling all dependence on the RG scale at leading-log accuracy. We employ this same “all-FCC-ee” likelihood for all results presented in this letter.

Our results are displayed in \cref{fig:Higgs} in the $(M_A,\tan\beta)$ plane, the only two independent parameters affecting our observables.  We limit $1< \tan\beta <60$ as dictated by the usual perturbativity reasons \cite{Martin:1997ns}.  The left and right panels correspond to the most conservative theory-limited (S1) and experimental-only (S3) FCC-ee benchmarks defined in \cref{app:EWPO}, respectively.  The 95\% CL contours include the oblique parameters $S, T, W, Y$ (dominated by $T$), shifts in Higgs couplings (dominated by $\kappa_{b,\tau}$), and deviations in $Z \to b \bar b$ (dominated by $R_b$), together with the requirement $\tan\beta\gtrsim5$ to reproduce the Higgs mass.\footnote{This amounts to saturating the tree-level upper bound on the Higgs mass within 10\%, which in natural SUSY is anyway required even including radiative corrections \cite{Draper:2011aa}.} 
On the right panel, we also added fine-tuning contours associated to the MSSM parameters $b, m_{H_u}^2, m_{H_d}^2$, as defined in~\cite{Perelstein:2007nx}.

\begin{figure*}[!htb]
    \centering
    \begin{subfigure}[t]{0.48\textwidth}
        \centering
        \includegraphics[width=\textwidth]{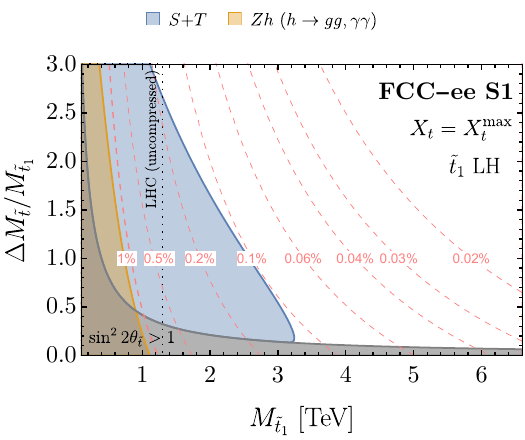} 
    \end{subfigure}
    \hfill
    \begin{subfigure}[t]{0.475\textwidth}
        \centering
        \includegraphics[width=\linewidth]{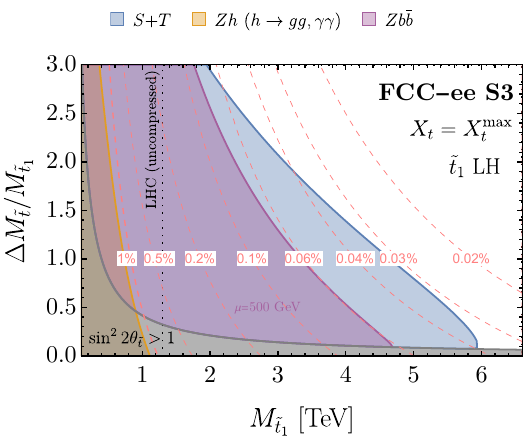} 
    \end{subfigure}
    \begin{subfigure}[t]{0.48\textwidth}
        \centering
        \includegraphics[width=\linewidth]{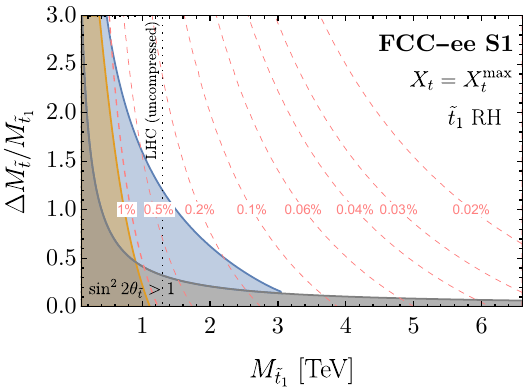} 
    \end{subfigure}
    \hfill
    \begin{subfigure}[t]{0.475\textwidth}
        \centering
        \includegraphics[width=\linewidth]{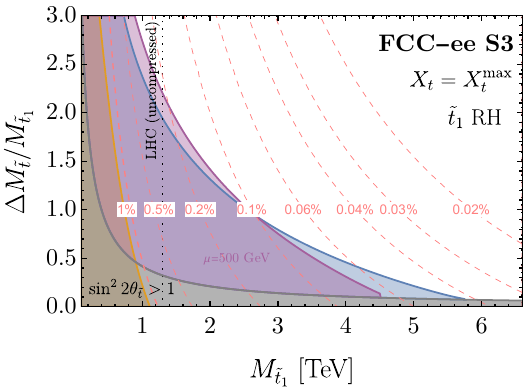} 
    \end{subfigure}
    \caption{\justifying FCC-ee projected $2\sigma$ sensitivities to the lightest stop mass $M_{\tilde t_1}$ in the mostly LH (top row) and RH (bottom row) cases, as function of the stop mass splitting $\Delta M_{\tilde t} / M_{\tilde t_1}$ and with the trilinear $X_t$ fixed to the value maximizing the Higgs mass, $X_t^{\rm max} = \sqrt{6M_{\tilde t_1}M_{\tilde t_2}}$. Shaded regions correspond to expected sensitivities from $S$ and $T$ (blue), Higgs decays (orange), and $Z\rightarrow b_L \bar b_L$ (purple). The black dotted line marks current LHC direct search limits in the uncompressed case~\cite{CMS:2021eha}. Fine-tuning contours are given by the dashed red lines. The gray region lies outside the physical parameter space, as $\left|\sin 2\theta_{\tilde t} \right| > 1$. See Section~\ref{sec:stops} for details.}
    \label{fig:DeltaPlotStop}
\end{figure*}

In the theory-limited S1 scenario (left panel), the dominant constraint is set by Higgs decays. The dependence on $\tan\beta$ is mild, yielding an almost flat bound of $M_A \gtrsim 1.7$~TeV. Shifts in the $Zb\bar b$ vertex provide a complementary bound to Higgs couplings for large $\tan \beta$. By comparison, direct LHC searches~\cite{ATLAS:2020zms, ATLAS:2024itc} constrain $M_A$ up to 2\,TeV for large $\tan \beta$, with weaker limits at low $\tan \beta$, as indicated by the black dotted line in \cref{fig:Higgs}. Unfortunately, in the S1 case, the reach of FCC-ee to probe the heavy Higgs of the MSSM is only complementary to limits set by current direct searches at the LHC. 

On the other hand, the S3 scenario, assuming only experimental uncertainties, highlights the remarkable potential of a fully exploited Tera-$Z$ run. If theoretical predictions can match experimental precision, we find that \emph{$R_b$ alone will dominate over all other observables}, including Higgs decays.
This yields a bound of $m_A \gtrsim 9$\,TeV for $\tan\beta \approx 50$, where the $m_t/m_b$ hierarchy is addressed and the theory is more natural as indicated by red-dashed lines. We find an overall lower bound of $m_A \gtrsim 2$\,TeV. To the best of our knowledge, this is the first time the dominance of $R_b$ in this scenario has been emphasized.

As a final point, let us comment on other bounds not included in our plots.
The HL-LHC is expected to improve the direct reach to $\sim 2.5$\,TeV~\cite{Baer:2025zqt} for large $\tan\beta$, surpassing the S1 scenario. 
In addition, the HL-LHC will further improve Higgs coupling measurements, enabling meaningful tests of the heavy Higgs sector at moderate $\tan\beta$, with sensitivity not far below that of FCC-ee.

\section{Stops} 
\label{sec:stops}

Scalar top partners (stops) are the prototypical supersymmetric particles and play a central role in the MSSM. They regulate the otherwise quadratically divergent radiative corrections to the electroweak scale. The large top Yukawa coupling also allows stop loops to lift the tree-level Higgs mass prediction of $M_h \leq M_Z$ through radiative corrections. This contribution grows logarithmically with the stop masses and is maximized for large mixing. In particular, even in the maximal mixing scenario, $X_t^{\rm max} = \sqrt{6M_{\tilde t_1}M_{\tilde t_2}}$~\cite{Haber:1996fp}, achieving the observed Higgs mass requires stops with mass $\gtrsim 1$\,TeV~\cite{PardoVega:2015eno}. Given the observed Higgs mass, it is unsurprising that LHC direct searches have not yet found evidence of stops. As we show here, however, FCC-ee offers the potential to probe well into this highly motivated parameter space.

Larger stop masses inevitably introduce greater fine-tuning to account for the little hierarchy between the soft SUSY-breaking masses and the electroweak scale. This is because the one-loop corrections to $m_{H_u}^2$, which directly determine $m_Z^2$~\cite{Martin:1997ns}, grow quadratically with the stop masses~\cite{Perelstein:2007nx}. 
Due to the large top Yukawa coupling and color factors, these loop contributions can rival tree-level terms and are typically the primary source of fine-tuning in the MSSM. Tightening experimental bounds on stops further increases the pressure on naturalness.

Given these important features, it is essential to evaluate FCC-ee sensitivity to stops.  The most important set of observables associated with these particles are: $i)$ oblique parameters $S$ and $T$, $ii)$ Higgs couplings to gluons and photons, and $iii)$ $R_b$ at the $Z$-pole. The first two receive 1-loop corrections involving only stops that, crucially, depend on just a handful of parameters: the mass eigenstates $M_{\tilde t_1}, M_{\tilde t_2}$ and the trilinear coupling $X_t$ entering via the stop mixing angle $\sin 2\theta_{\tilde t} = 2 M_t X_t/(M_{\tilde t_2}
^2 - M_{\tilde t_1}^2)$ \cite{Martin:1997ns}.  
Their expressions can be found in~\cite{Arvanitaki:2011ck, Fan:2014axa}. Here, we neglect bottom mixing and note that the dependence on $\tan\beta$ is mild, effectively vanishing for $\tan\beta \gtrsim 5$. 
On the other hand, the dominant corrections to $R_b$ here arise from mixed stop-Higgsino loops, which also depend on the Higgsino mass~\cite{deBoer:1996hd}. 

Once again, we present two sets of plots showing the 95\% CL limits from these observables for the theory-limited S1 scenario (left column) and the experimental-only S3 scenario (right column) in~\cref{fig:DeltaPlotStop}. In these plots, we fix $X_t = X_t^{\text{max}}$ to maximize the radiative corrections to the Higgs mass and avoid regions where the Higgs mass would be too light. We define $M_{\tilde t_1}$ as the mass of the lightest stop eigenstate, so that the vertical axis shows the mass splitting $\Delta M_{\tilde t} = M_{\tilde t_2} - M_{\tilde t_1}$, normalized to $M_{\tilde t_1}$. In motivated scenarios, this ratio is expected to be $\lesssim 1$, though not exactly degenerate, which would introduce additional tuning (for top-down predictions, see e.g. Ref.~\cite{Antusch:2015nwi}).The plots distinguish the cases where $\tilde{t}_1$ is predominantly left-handed (top row) or right-handed (bottom row), corresponding to the two physically distinct solutions for the angle given by $\sin 2\theta_{\tilde{t}} = 2m_t X_t/(M_{\tilde{t}_2}^2 - M_{\tilde{t}_1}^2)$.\footnote{RG effects tend to decrease the RH stop mass parameter more, which could prefer a scenario where the lightest stop is mostly RH \cite{Martin:1997ns}.}

In the conservative scenario S1, the strongest probes of stops come from the $S$ and $T$ parameters, where $T$ dominates the bound. Here, electroweak precision tests again provide the leading sensitivity, surpassing Higgs decays and setting a bound of $M_{\tilde t_{1}} \gtrsim 2.5$ (1.5) TeV in the LH (RH) cases for $\mathcal{O}(1)$ mass splittings. Not only does this outperform direct LHC reach for uncompressed spectra~\cite{CMS:2021eha} (dotted black), but it also closes the possibility of light stops in the compressed limit. As shown by the red-dashed fine-tuning contours associated with $m_{H_u}^2$, FCC-ee can test naturalness in the context of the MSSM at the per-mille level. In the experimental-only S3 scenario, $T$ probes LH stops up to 5 TeV for $\Delta M_{\tilde t}/ M_{\tilde t_1} \sim 1$, corresponding to fine-tuning at the $10^{-4}$ level. Compared to current LHC constraints, this is an improvement on fine-tuning by more than an order of magnitude. 
We stress that for fixed $X_t$, the $T$ parameter bound can only be evaded by decoupling at least one of the stops, which comes at the expense of increased fine-tuning.

The extreme precision on $R_b$ in the experimental-only S3 case renders it once again a relevant observable, also for stops. Here we employ an EFT approach that is valid in the limit of small stop mixing. Note that this does not necessarily require small $X_t$, but rather a separation from the gray-shaded region in~\cref{fig:DeltaPlotStop} where $\sin 2\theta_{\tilde t}$ becomes $O(1)$ and the EFT breaks down. Contrary to the heavy Higgs, the dominant contributions to $R_b$ arise from corrections to the $Z$ vertex with left-handed bottoms, generated at 1-loop from mixed graphs involving stops and Higgsinos. We provide the full 1-loop matching in \cref{app:WCs}.\footnote{Our results agree with~\cite{Kraml:2025fpv} but disagree with $\Delta R_b$ given in Appendix A.3 of \cite{Fan:2014axa} when the small-angle (EFT) limit is taken.} Importantly, we have $\delta g_L^{Zb} \propto [C_{Hq}^{(1)}]_{33} + [C_{Hq}^{(3)}]_{33}$, so only the terms scaling as $y_t^2 g^{(\prime)2}$ and $y_t^4 X_t^2$ contribute to $R_b$. 
In natural SUSY scenarios, Higgsinos are expected to be light due to their tree-level contribution to electroweak fine-tuning. Interestingly, light Higgsinos also result in the largest contributions to $R_b$.

In \cref{fig:DeltaPlotStop} right, the purple region shows the sensitivity of $R_b$ for $\tan\beta \gtrsim 10$ and fixing the Higgsino mass to $\mu =500$\,GeV, consistent with the limits we obtain in~\cref{sec:gaugino}. For light Higgsinos, $R_b$ depends on the Higgsino mass only logarithmically, so the sensitivity does not change much except in the unnatural limit of $\mu \gtrsim M_{\tilde t_1}$.
We find that $R_b$ offers a complementary probe (especially in the mostly RH case) where correlated signals may appear in both the $T$ parameter and shifts to the $Zb_L \bar{b}_L$ vertex. The latter can be disentangled from right-handed corrections using a combination of $R_b$ and $A_{\rm FB}^b$, making both observables important targets for further theoretical development.
Finally, as we show  in~\cref{app:plots}, $R_b$ can become the dominant probe for some of the parameter space where $X_t < X_t^{\rm max}$.

\section{Gauginos and Higgsinos} 
\label{sec:gaugino}

As a final benchmark, we consider Higgsinos and electroweak gauginos. Along with gluinos, these are the only additional fermionic states predicted by the MSSM and can naturally serve as dark matter candidates if the corresponding neutralino is the lightest supersymmetric particle. Moreover, the Higgsino mass parameter $\mu$ contributes to electroweak fine‐tuning at tree level \cite{Perelstein:2007nx}, so any lower bound on $\mu$ directly translates into a statement about naturalness. These states also serve as canonical benchmarks for a future $e^+e^-$ linear collider facility (LCF), where they could be pair‐produced in a clean environment. Thus, they offer a useful benchmark for comparing the reach of linear and circular $e^+e^-$ colliders.

\begin{figure}[!htb]
    \centering
    \hspace{-10mm}
    \includegraphics[scale=0.95]{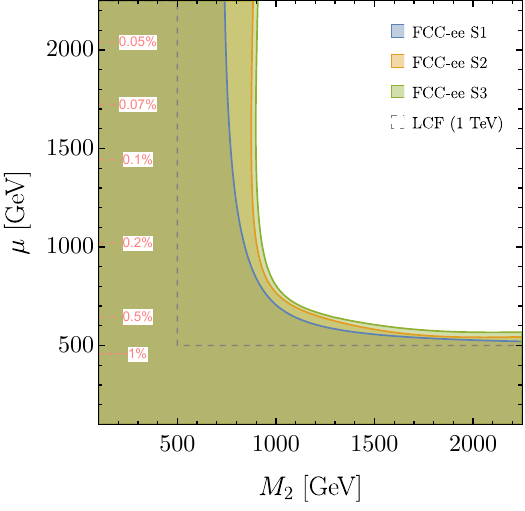}
    \caption{\justifying FCC-ee projected $2\sigma$ sensitivities to the Higgsino and Gaugino parameters $\mu$ and $M_2$ arising from the oblique parameters $S,T,W,Y$. The three contours represent the theory scenarios outlined in \cref{app:EWPO}. The dashed gray line represents the ``na\"ive'' projection for direct searches at a 1 TeV LCF. Dashed red lines give the fine-tuning associated with $\mu$. See Section~\ref{sec:gaugino} for details.}
    \label{fig:higgsino}
\end{figure}

The FCC-ee program constrains the Higgsino mass $\mu$ and the bino/wino masses $M_1$ and $M_2$ through precision measurements of the universal oblique parameters $S, T, W, Y$ at the Tera-$Z$ run, as well as fermion pair production above the $Z$-pole. These depend only on the three masses $\mu$, $M_1$, and $M_2$. The $S$ and $T$ parameters also have a mild dependence on $\tan\beta$ due to mixing after electroweak symmetry breaking. To obtain the associated bounds, we use an EFT approach where these states are integrated out and matched to the SMEFT at 1-loop, reproducing the results of~\cite{Marandella:2005wc}. This approach is justified a posteriori since the mass limits we obtain always lie above the scale at which the corresponding measurements are performed. 

In \cref{fig:higgsino} we display the 95\% CL constraints in the $(M_2,\mu)$ plane,  where we include all the three scenarios provided in~\cref{app:EWPO}. Remarkably, we find that the limits are entirely driven by the $W$ parameter, $\hat{W} = \alpha_L m_W^2/(30\pi) (1/\mu^{2} + 2/ M_2^{2})$. The bound does not vary much in the three scenarios, as $W$ is best constrained by fermion pair-production above the $Z$-pole where theory errors are subleading~\cite{Greljo:2024ytg}. Furthermore, the choice of $M_1$ and $\tan\beta$ has negligible impact on our results since only $S,T$ depend on them and their contributions are subdominant.\footnote{For concreteness, we impose the GUT‐inspired relation $M_1 = M_2/2$ and fix $\tan\beta=10$.}
It is important to note that any other MSSM contributions to $W$ add constructively~\cite{Marandella:2005wc}, so our results represent the \emph{absolutely lowest allowed values for $\mu$ and $M_2$}. The weakest constraint is on $\mu$, which still must exceed roughly 500\,GeV. Interestingly, this matches and slightly surpasses the na\"ive direct reach at a 1\,TeV LCF.

Finally, we note that these indirect limits are decay-channel independent and remain valid even for compressed spectra. In contrast, LHC direct searches currently exclude charginos and neutralinos up to $\lesssim 1$\,TeV but lose sensitivity when mass splittings are small~\cite{ATLAS:2024qxh}. Moreover, pure Higgsino dark matter scenarios are typically realized in a compressed regime. The exceptional precision of a Tera-$Z$ run would therefore offer invaluable, complementary coverage to close the gaps left by the LHC.

\section{Conclusions} 
\label{sec:conc}

The FCC program is primarily motivated by the quest to understand the microscopic origin of the Higgs boson. Among proposed theories, the MSSM remains a leading candidate. In this letter, we have studied the potential of FCC-ee to probe the natural parameter space of three key sectors of the MSSM: the heavy Higgs doublet, scalar top partners, and light gauginos and higgsinos. Contrary to na\"ive expectations that the Higgs factory mode of FCC-ee offers the strongest probes, we find that the Tera-$Z$ run provides superior sensitivity due to its immense statistical precision, allowing detection of subtle quantum effects. 

To realize the full potential of Tera-$Z$, a vigorous effort is needed to reduce SM theoretical uncertainties. We illustrate this by constructing three benchmark scenarios S1-S3 (\cref{app:EWPO}), where S1 is strongly theory limited while S3 considers only experimental errors to explore the ultimate reach of the Tera-$Z$ program. We also identify electroweak precision observables that probe non-oblique shifts of the $Zb\bar b$ vertex, in particular $R_b$, as areas of maximum impact for a focused theoretical effort in the context of the MSSM.

Overall, the FCC program presents an exceptional opportunity for MSSM exploration. Indirect reach at FCC-ee is already impressive, probing the MSSM in the multi-TeV range, closing compressed spectra blind spots at the LHC, and testing the naturalness of the EW scale at least at the per-mille level. Longer term, direct searches for sparticles at FCC-hh will surpass the indirect reach of FCC-ee, pushing LHC limits by an order of magnitude and probing electroweak fine tuning at the $10^{-4}$ level \cite{FCC:2025lpp}.

\section*{Acknowledgements}

We thank Stefan Antusch, Nathaniel Craig, Howard Haber, and Sophie Renner for helpful discussions. We also thank Sophie Renner for her help in constructing the S1-S3 benchmarks in~\cref{app:EWPO}. The work of AG and AV has received funding from
the Swiss National Science Foundation (SNF) through the
Eccellenza Professorial Fellowship “Flavor Physics at the
High Energy Frontier,” project number 186866. The work of BAS is supported by a CDEIGENT grant from the Generalitat Valenciana with grant no. CIDEIG/2023/35. The authors are grateful to the Mainz Institute for Theoretical Physics (MITP) of the Cluster of Excellence PRISMA+ (Project ID 390831469), for its hospitality and support during the completion of this work. 

\appendix
\section{Updated Electroweak Precision Observables}
\label{app:EWPO}
As part of the ESPP update, the Electroweak Physics Preparatory Group (EW PPG) has recently presented results for theoretical uncertainties on electroweak precision observables for future $e^+ e^-$ colliders~\cite{EWPPGTH}. Based on these results, we have constructed three benchmark scenarios: S1 (conservative theory progress), S2 (aggressive theory progress), and S3 (experimental errors only). More specifically, there are two categories of theoretical errors outlined by the EW PPG: i) errors associated to the conversion of experimental measurements to the electroweak ``pseudo-observables" and ii) uncertainties on SM theory predictions (e.g. due to missing higher orders). The EW PPG has presented one scenario for errors of type i) that we label (TH PO), while they provide conservative and aggressive scenarios for errors of type ii) that we label (TH cons.) and (TH agg.). The advised prescription was to add theory errors of type i)+ii) in quadrature with experimental statistical and systematic uncertainties, so we construct three scenarios as follows
\begin{itemize}
\item S1: TH PO+TH agg.+EXP
\item S2: TH agg.+EXP
\item S3: EXP Only ,
\end{itemize}
where ``+" indicates that the errors are combined in quadrature. Here, (EXP) refers to experimental statistical and systematic uncertainties that we take from Refs.~\cite{Selvaggi:2025kmd,FCC:2025lpp} added in quadrature. We have opted not to include a scenario including the TH cons. scenario for type ii) errors since TH PO errors of type i) tend to be comparable or larger for most observables. Thus, the overall philosophy is that S1 corresponds to a scenario where type i) errors dominate, while S2 is a case where a lot of theory progress has been made and type ii) errors dominate (but have not quite reached experimental precision), while S3 is the ultimate scenario showing pure experimental reach.

For observables directly provided by the EW PPG, we use the numbers given in Ref.~\cite{EWPPGTH}. For the rest, we obtain error estimates by expressing the unknown observables in terms of the known ones and using na\"ive error propagation. The final results of this exercise are given in~\cref{table:EWPOupdate}, where all numbers are all relative uncertainties with an overall $10^{-5}$ factored out.

\begin{table*}[ht]
\centering
\begin{tabular}{|l|c|c|c|}
\hline
& \textbf{Scenario S1} & \textbf{Scenario S2} & \textbf{Scenario S3} \\
\hline\hline
\textbf{Observable} & \textbf{TH PO+TH agg.+EXP ($10^{-5}$)} & \textbf{TH agg.+EXP ($10^{-5}$)} & \textbf{EXP Only ($10^{-5}$)} \\
\hline
$\Gamma_Z$ & 1.55 & 0.820 & 0.510 \\
$\sigma_{\text{had}}$ & 4.33 & 2.06 & 1.93 \\
$R_e$ & 2.21 & 1.05 & 0.410 \\
$R_\mu$ & 2.20 & 1.02 & 0.330 \\
$R_\tau$ & 2.20 & 1.03 & 0.350 \\
$R_b$ & 20.1 & 1.63 & 0.180 \\
$R_c$ & 100 & 1.19 & 0.260 \\
$A_{\text{FB}}^e$ & 126 & 25.7 & 25.2 \\
$A_{\text{FB}}^\mu$ & 125 & 21.1 & 20.6 \\
$A_{\text{FB}}^\tau$ & 126 & 23.3 & 22.8 \\
$A_{\text{FB}}^b$ & 87.8 & 6.42 & 5.50 \\
$A_{\text{FB}}^c$ & 89.1 & 10.2 & 9.62 \\
$A_{\text{FB}}^s$ & 88.2 & 10.7 & 10.2 \\
$\sin^2\theta_W$ & 6.87 & 0.780 & 0.730 \\
$A_e$ & 87.9 & 9.78 & 9.20 \\
$A_\mu$ & 90.1 & 22.1 & 21.8 \\
$A_\tau$ & 90.5 & 23.4 & 23.2 \\
$A_b$ & 11.7 & 10.5 & 10.5 \\
$A_c$ & 16.9 & 9.00 & 8.99 \\
$A_s$ & 14.2 & 13.2 & 13.2 \\
$M_W$ & 0.490 & 0.320 & 0.300 \\
$\Gamma_W$ & 16.1 & 16.1 & 16.1 \\
\hline
\end{tabular}
\caption{Relative uncertainties for $Z$ and $W$ pole observables. Three different scenarios for theoretical uncertainties (derived from~\cite{EWPPGTH}) are considered. Experimental numbers are taken from~\cite{Selvaggi:2025kmd,FCC:2025lpp}. See text for more details.}
\label{table:EWPOupdate}
\end{table*}

\section{S,T,W,Y fit}
\label{app:STWY}
Using our ``all-FCC-ee" likelihood that includes electroweak precision tests at the $Z$- and $W$-poles as well as fermion pair-production ratios above the $Z$-pole, we perform a fit to the oblique parameters $\hat S,\hat T,\hat W,\hat Y$~\cite{Barbieri:2004qk} for the three scenarios given in~\cref{table:EWPOupdate}. We report the Gaussian likelihoods here at the scale $M_Z$ (neglecting RGE for the datasets at 163, 240, 365 GeV).

\subsection{FCC-ee S1 Scenario}
The $\hat S,\hat T,\hat W,\hat Y$ fit in the S1 case yields
\begin{equation}
\left(
\begin{array}{cc}
\hat{S}  \\
\hat{T} \\
\hat{W} \\
\hat{Y} \\
\end{array}
\right)= \pm
\left(
\begin{array}{cc}
2.42 \\
1.29 \\
0.48 \\
1.55 \\
\end{array}
\right)\times 10^{-5}\,,
\label{eq:STWY}
\end{equation}
and their correlation matrix is
\begin{equation}
\rho = \left(
\begin{array}{cccc}
 1. & 0.856 & 0.315 & 0.672 \\
 0.856 & 1. & 0.154 & 0.352 \\
 0.315 & 0.154 & 1. & 0.221 \\
 0.672 & 0.352 & 0.221 & 1. \\
\end{array}
\right)\,.
\label{eq:STWYcorr}
\end{equation}

\subsection{FCC-ee S2 Scenario}
The $\hat S,\hat T,\hat W,\hat Y$ fit in the S2 case yields
\begin{equation}
\left(
\begin{array}{cc}
\hat{S}  \\
\hat{T} \\
\hat{W} \\
\hat{Y} \\
\end{array}
\right)= \pm
\left(
\begin{array}{cc}
1.74 \\
0.73 \\
0.47 \\
1.55 \\
\end{array}
\right)\times 10^{-5}\,,
\label{eq:STWY}
\end{equation}
and their correlation matrix is
\begin{equation}
\rho = \left(
\begin{array}{cccc}
 1. & 0.825 & 0.394 & 0.937 \\
 0.825 & 1. & 0.248 & 0.63 \\
 0.394 & 0.248 & 1. & 0.216 \\
 0.937 & 0.63 & 0.216 & 1. \\
\end{array}
\right)\,.
\label{eq:STWYcorr}
\end{equation}

\subsection{FCC-ee S3 Scenario}
The $\hat S,\hat T,\hat W,\hat Y$ fit in the S3 case yields
\begin{equation}
\left(
\begin{array}{cc}
\hat{S}  \\
\hat{T} \\
\hat{W} \\
\hat{Y} \\
\end{array}
\right)= \pm
\left(
\begin{array}{cc}
1.71 \\
0.63 \\
0.46 \\
1.55 \\
\end{array}
\right)\times 10^{-5}\,,
\label{eq:STWY}
\end{equation}
and their correlation matrix is
\begin{equation}
\rho = \left(
\begin{array}{cccc}
 1. & 0.885 & 0.414 & 0.952 \\
 0.885 & 1. & 0.365 & 0.749 \\
 0.414 & 0.365 & 1. & 0.211 \\
 0.952 & 0.749 & 0.211 & 1. \\
\end{array}
\right)\,.
\label{eq:STWYcorr}
\end{equation}

\section{Wilson coefficients}
\label{app:WCs}

\paragraph{Higgsino-stop 1-loop matching.} The WCs of the SMEFT operators $[O_{Hq}^{(1),(3)}]_{33}$ generated by Higgsino-stop loops read
\begin{widetext}
\begin{align}
&
\begin{aligned}
\bigl[C_{Hq}^{(1)}\bigr]_{33} &=
\frac{y_t^2}{6912\pi^2\,M_{U_3}^2\,\sin^2\beta}
\bigg\{
\frac{g'^2}{(1 - x)^4}\Big[
(1 - x)(56-115x+65x^2)
+ 6\,(6 - 9x + 4x^3) \log x
\Big]\\
&\quad
+\,\frac{36y_t^2}{(r - x)^4}\Big[
(r - x)(r^2 - 5rx - 2x^2)
+ 6\,r\,x^2\log \frac{r}{x}
\Big]
\bigg\} \\[8pt]
&\quad
+\;\frac{y_t^4\,X_t^2}{64\pi^2\,M_{U_3}^4\,\sin^2\beta}
\;\frac{1}{(r-1)^3(r - x)^2(1 - x)^2}
\Big[
(1 - x)(r-1)(r - x)\bigl(2r-rx - x\bigr)\\
&\quad
- r(r^2 + r - 2x)(1-x)^2\,\log r
+(r - 1)^3 x^2 \log x
\Big]
\end{aligned}
\\
&
\begin{aligned}
\bigl[C_{Hq}^{(3)} \bigr]_{33}&=
-\frac{y_t^2}{6912\pi^2\,M_{U_3}^2\,\sin^2\beta} 
\Bigg\{
\frac{3 g^2}{(1 - x)^4}\Big[(1-x)(16 - 29x + 7x^2) 
 + 6\,(2 - 3x)\log x
\Big] \\
&\quad +\frac{36y_t^2}{(r-x)^4}\Big[
\,(r - x)(r^2 - 5r x - 2x^2)
+ 6\,r\, x^2\,\log\frac{r}{x}
\Big]
\Bigg\}
\end{aligned}
\end{align}
\end{widetext}
where $x=\mu^2/M_{U_3}^2$ and $r= M_{Q_3}^2/M_{U_3}^2$. 

\section{Additional plots}

Here we append additional plots complementing Section~\cref{sec:stops}.

\label{app:plots}
\begin{figure*}[!htb]
    \centering
    \includegraphics[scale=1]{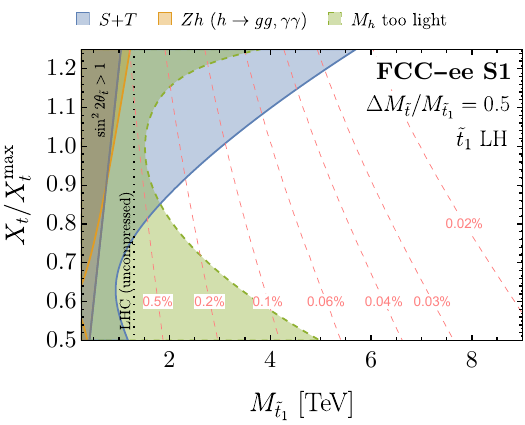} \hspace{3mm}
    \includegraphics[scale=1]{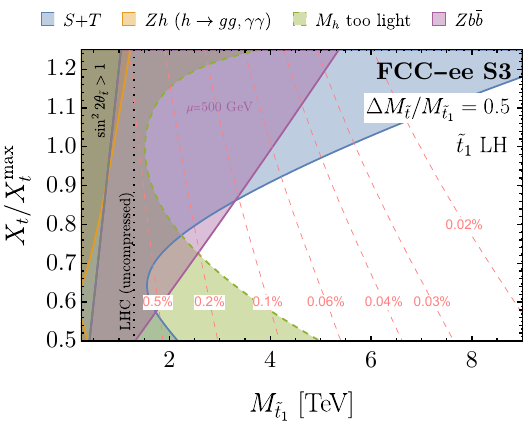}
    \caption{\justifying FCC-ee projected $2\sigma$ sensitivities to the lightest stop mass (assuming it is mostly LH) as function of $X_t$ for fixed mass splitting $\Delta M_{\tilde t}/M_{\tilde t_1} = 0.5$. Shaded regions correspond to expected sensitivities from $S$ and $T$ (blue), Higgs decays (orange), and $Z\rightarrow b_L \bar b_L$ (purple). The Higgs mass constraint region (green) is obtained using SUSYHD~\cite{PardoVega:2015eno}. The black dotted line marks current LHC direct search limits in the uncompressed case~\cite{CMS:2021eha}. Fine-tuning contours are given by the dashed red lines. The gray region lies outside the physical parameter space, as $\left|\sin 2\theta_{\tilde t} \right| > 1$. See Section~\ref{sec:stops} for details.}
    \label{fig:XtPlotStop}
\end{figure*}

\begin{figure*}[!htb]
    \centering
    \includegraphics[scale=1]{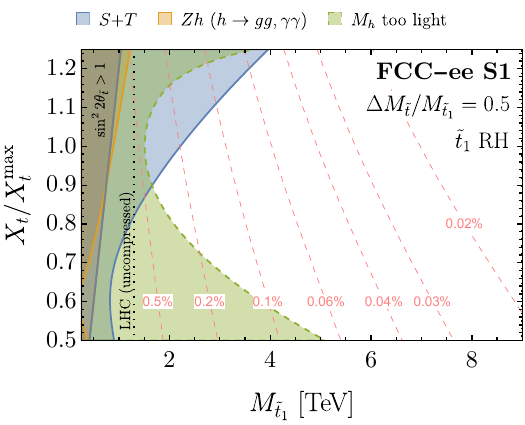} \hspace{3mm}
    \includegraphics[scale=1]{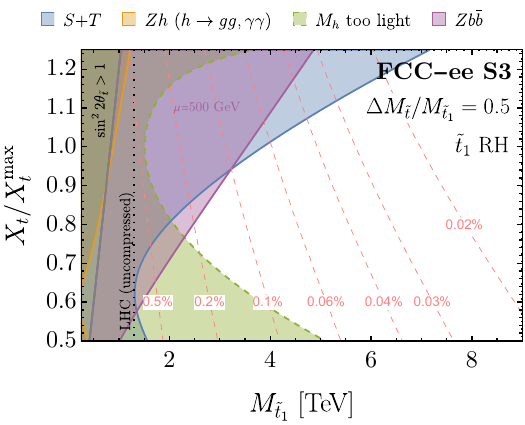}
    \caption{\justifying FCC-ee projected $2\sigma$ sensitivities to the lightest stop mass (assuming it is mostly RH) as function of $X_t$ for fixed mass splitting $\Delta M_{\tilde t}/M_{\tilde t_1} = 0.5$. Shaded regions correspond to expected sensitivities from $S$ and $T$ (blue), Higgs decays (orange), and $Z\rightarrow b_L \bar b_L$ (purple). The Higgs mass constraint region (green) is obtained using SUSYHD~\cite{PardoVega:2015eno}. The black dotted line marks current LHC direct search limits in the uncompressed case~\cite{CMS:2021eha}. Fine-tuning contours are given by the dashed red lines. The gray region lies outside the physical parameter space, as $\left|\sin 2\theta_{\tilde t} \right| > 1$. See Section~\ref{sec:stops} for details.}
    \label{fig:XtPlotStopRH}
\end{figure*}

\bibliographystyle{JHEP}
\bibliography{bibliography.bib}

\providecommand{\href}[2]{#2}\begingroup\raggedright\begin{thebibliography}{10}

\bibitem{Haber:1984rc}
H.~E. Haber and G.~L. Kane, {\it {The Search for Supersymmetry: Probing Physics
  Beyond the Standard Model}},  {\em Phys. Rept.} {\bf 117} (1985) 75--263.

\bibitem{Martin:1997ns}
S.~P. Martin, {\it {A Supersymmetry primer}},  {\em Adv. Ser. Direct. High
  Energy Phys.} {\bf 18} (1998) 1--98,
  [\href{http://arxiv.org/abs/hep-ph/9709356}{{\tt hep-ph/9709356}}].

\bibitem{Dreiner:2023yus}
H.~K. Dreiner, H.~E. Haber, and S.~P. Martin, {\em {From Spinors to
  Supersymmetry}}.
\newblock Cambridge University Press, Cambridge, UK, 7, 2023.

\bibitem{Allanach:2024suz}
B.~Allanach and H.~E. Haber, {\it {Supersymmetry, Part I (Theory)}},
  \href{http://arxiv.org/abs/2401.03827}{{\tt arXiv:2401.03827}}.

\bibitem{EuropeanStrategyforParticlePhysicsPreparatoryGroup:2019qin}
R.~K. Ellis et~al., {\it {Physics Briefing Book}: {Input for the European
  Strategy for Particle Physics Update 2020}},
  \href{http://arxiv.org/abs/1910.11775}{{\tt arXiv:1910.11775}}.

\bibitem{Altmann:2025feg}
J.~Altmann et~al., {\em {ECFA Higgs, electroweak, and top Factory Study}},
  vol.~5/2025 of {\em CERN Yellow Reports: Monographs}.
\newblock 6, 2025.

\bibitem{FCC:2025lpp}
{\bf FCC} Collaboration, M.~Benedikt et~al., {\it {Future Circular Collider
  Feasibility Study Report: Volume 1, Physics, Experiments, Detectors}},
  \href{http://arxiv.org/abs/2505.00272}{{\tt arXiv:2505.00272}}.

\bibitem{CEPCStudyGroup:2018ghi}
{\bf CEPC Study Group} Collaboration, M.~Dong et~al., {\it {CEPC Conceptual
  Design Report: Volume 2 - Physics \& Detector}},
  \href{http://arxiv.org/abs/1811.10545}{{\tt arXiv:1811.10545}}.

\bibitem{Anastopoulos:2025jyh}
C.~Anastopoulos et~al., {\it {LEP3: A High-Luminosity e+e- Higgs and
  ElectroweakFactory in the LHC Tunnel}},
  \href{http://arxiv.org/abs/2504.00541}{{\tt arXiv:2504.00541}}.

\bibitem{ILC:2013jhg}
{\bf ILC} Collaboration, {\it {The International Linear Collider Technical
  Design Report - Volume 2: Physics}},
  \href{http://arxiv.org/abs/1306.6352}{{\tt arXiv:1306.6352}}.

\bibitem{CLICdp:2018cto}
{\bf CLICdp, CLIC} Collaboration, T.~K. Charles et~al., {\it {The Compact
  Linear Collider (CLIC) - 2018 Summary Report}},  {\em CERN Yellow Rep.
  Monogr.} {\bf 2} (2018) 1--112, [\href{http://arxiv.org/abs/1812.06018}{{\tt
  arXiv:1812.06018}}].

\bibitem{LinearCollider:2025lya}
{\bf Linear Collider} Collaboration, A.~Subba et~al., {\it {The Linear Collider
  Facility (LCF) at CERN}},  \href{http://arxiv.org/abs/2503.24049}{{\tt
  arXiv:2503.24049}}.

\bibitem{Allwicher:2023shc}
L.~Allwicher, C.~Cornella, G.~Isidori, and B.~A. Stefanek, {\it {New physics in
  the third generation. A comprehensive SMEFT analysis and future prospects}},
  {\em JHEP} {\bf 03} (2024) 049, [\href{http://arxiv.org/abs/2311.00020}{{\tt
  arXiv:2311.00020}}].

\bibitem{Allwicher:2024sso}
L.~Allwicher, M.~McCullough, and S.~Renner, {\it {New physics at Tera-Z:
  precision renormalised}},  {\em JHEP} {\bf 02} (2025) 164,
  [\href{http://arxiv.org/abs/2408.03992}{{\tt arXiv:2408.03992}}].

\bibitem{Greljo:2024ytg}
A.~Greljo, H.~Tiblom, and A.~Valenti, {\it {New Physics Through Flavor Tagging
  at FCC-ee}},  \href{http://arxiv.org/abs/2411.02485}{{\tt arXiv:2411.02485}}.

\bibitem{Maura:2024zxz}
V.~Maura, B.~A. Stefanek, and T.~You, {\it {Accuracy complements energy:
  electroweak precision tests at Tera-Z}},
  \href{http://arxiv.org/abs/2412.14241}{{\tt arXiv:2412.14241}}.

\bibitem{Stefanek:2024kds}
B.~A. Stefanek, {\it {Non-universal probes of composite Higgs models: new
  bounds and prospects for FCC-ee}},  {\em JHEP} {\bf 09} (2024) 103,
  [\href{http://arxiv.org/abs/2407.09593}{{\tt arXiv:2407.09593}}].

\bibitem{Erdelyi:2024sls}
B.~A. Erdelyi, R.~Gr{\"o}ber, and N.~Selimovic, {\it {How large can the light
  quark Yukawa couplings be?}},  {\em JHEP} {\bf 05} (2025) 189,
  [\href{http://arxiv.org/abs/2410.08272}{{\tt arXiv:2410.08272}}].

\bibitem{Gargalionis:2024jaw}
J.~Gargalionis, J.~Quevillon, P.~N.~H. Vuong, and T.~You, {\it {Linear Standard
  Model extensions in the SMEFT at one loop and Tera-Z}},
  \href{http://arxiv.org/abs/2412.01759}{{\tt arXiv:2412.01759}}.

\bibitem{Celada:2024mcf}
E.~Celada, T.~Giani, J.~ter Hoeve, L.~Mantani, J.~Rojo, A.~N. Rossia, M.~O.~A.
  Thomas, and E.~Vryonidou, {\it {Mapping the SMEFT at high-energy colliders:
  from LEP and the (HL-)LHC to the FCC-ee}},  {\em JHEP} {\bf 09} (2024) 091,
  [\href{http://arxiv.org/abs/2404.12809}{{\tt arXiv:2404.12809}}].

\bibitem{Davighi:2024syj}
J.~Davighi, {\it {In Search of an Invisible $Z^\prime$}},
  \href{http://arxiv.org/abs/2412.07694}{{\tt arXiv:2412.07694}}.

\bibitem{Erdelyi:2025axy}
B.~A. Erdelyi, R.~Gr{\"o}ber, and N.~Selimovic, {\it {Probing new physics with
  the electron Yukawa coupling}},  {\em JHEP} {\bf 05} (2025) 135,
  [\href{http://arxiv.org/abs/2501.07628}{{\tt arXiv:2501.07628}}].

\bibitem{Maura:2025rcv}
V.~Maura, B.~A. Stefanek, and T.~You, {\it {The Higgs Self-Coupling at
  FCC-ee}},  \href{http://arxiv.org/abs/2503.13719}{{\tt arXiv:2503.13719}}.

\bibitem{Bordone:2025cde}
M.~Bordone, C.~Cornella, and J.~Davighi, {\it {Precision Tests in $b\to
  s\ell^+\ell^-$ ($\ell=e,\mu$) at FCC-ee}},
  \href{http://arxiv.org/abs/2503.22635}{{\tt arXiv:2503.22635}}.

\bibitem{terHoeve:2025gey}
J.~ter Hoeve, L.~Mantani, J.~Rojo, A.~N. Rossia, and E.~Vryonidou, {\it
  {Connecting scales: RGE effects in the SMEFT at the LHC and future
  colliders}},  {\em JHEP} {\bf 06} (2025) 125,
  [\href{http://arxiv.org/abs/2502.20453}{{\tt arXiv:2502.20453}}].

\bibitem{Allwicher:2025bub}
L.~Allwicher, G.~Isidori, and M.~Pesut, {\it {Flavored circular collider:
  cornering New Physics at FCC-ee via flavor-changing processes}},  {\em Eur.
  Phys. J. C} {\bf 85} (2025), no.~6 631,
  [\href{http://arxiv.org/abs/2503.17019}{{\tt arXiv:2503.17019}}].

\bibitem{deBlas:2021jlt}
J.~de~Blas, {\it {New physics at the FCC-ee: indirect discovery potential}},
  {\em Eur. Phys. J. Plus} {\bf 136} (2021), no.~9 897.

\bibitem{deBlas:2022ofj}
J.~de~Blas, Y.~Du, C.~Grojean, J.~Gu, V.~Miralles, M.~E. Peskin, J.~Tian,
  M.~Vos, and E.~Vryonidou, {\it {Global SMEFT Fits at Future Colliders}},  in
  {\em {Snowmass 2021}}, 6, 2022.
\newblock \href{http://arxiv.org/abs/2206.08326}{{\tt arXiv:2206.08326}}.

\bibitem{Azzi:2025dwl}
{\bf FCC} Collaboration, {\it {Prospects in BSM physics at FCC}}, .

\bibitem{Fan:2014vta}
J.~Fan, M.~Reece, and L.-T. Wang, {\it {Possible Futures of Electroweak
  Precision: ILC, FCC-ee, and CEPC}},  {\em JHEP} {\bf 09} (2015) 196,
  [\href{http://arxiv.org/abs/1411.1054}{{\tt arXiv:1411.1054}}].

\bibitem{Fan:2014axa}
J.~Fan, M.~Reece, and L.-T. Wang, {\it {Precision Natural SUSY at CEPC, FCC-ee,
  and ILC}},  {\em JHEP} {\bf 08} (2015) 152,
  [\href{http://arxiv.org/abs/1412.3107}{{\tt arXiv:1412.3107}}].

\bibitem{Knapen:2024bxw}
S.~Knapen, K.~Langhoff, and Z.~Ligeti, {\it {Imprints of supersymmetry at a
  future Z factory}},  {\em Phys. Rev. D} {\bf 111} (2025), no.~11 115007,
  [\href{http://arxiv.org/abs/2407.13815}{{\tt arXiv:2407.13815}}].

\bibitem{Nagata:2025ycf}
N.~Nagata and G.~Osaki, {\it {Electroweak Precision Data as a Gateway to Light
  Higgsinos}},  \href{http://arxiv.org/abs/2503.20602}{{\tt arXiv:2503.20602}}.

\bibitem{DAmbrosio:2002vsn}
G.~D'Ambrosio, G.~F. Giudice, G.~Isidori, and A.~Strumia, {\it {Minimal flavor
  violation: An Effective field theory approach}},  {\em Nucl. Phys. B} {\bf
  645} (2002) 155--187, [\href{http://arxiv.org/abs/hep-ph/0207036}{{\tt
  hep-ph/0207036}}].

\bibitem{Misiak:2015xwa}
M.~Misiak et~al., {\it {Updated NNLO QCD predictions for the weak radiative
  B-meson decays}},  {\em Phys. Rev. Lett.} {\bf 114} (2015), no.~22 221801,
  [\href{http://arxiv.org/abs/1503.01789}{{\tt arXiv:1503.01789}}].

\bibitem{Barbieri:1987fn}
R.~Barbieri and G.~F. Giudice, {\it {Upper Bounds on Supersymmetric Particle
  Masses}},  {\em Nucl. Phys. B} {\bf 306} (1988) 63--76.

\bibitem{ATLAS:2020zms}
{\bf ATLAS} Collaboration, G.~Aad et~al., {\it {Search for heavy Higgs bosons
  decaying into two tau leptons with the ATLAS detector using $pp$ collisions
  at $\sqrt{s}=13$ TeV}},  {\em Phys. Rev. Lett.} {\bf 125} (2020), no.~5
  051801, [\href{http://arxiv.org/abs/2002.12223}{{\tt arXiv:2002.12223}}].

\bibitem{deBlas:2017xtg}
J.~de~Blas, J.~C. Criado, M.~Perez-Victoria, and J.~Santiago, {\it {Effective
  description of general extensions of the Standard Model: the complete
  tree-level dictionary}},  {\em JHEP} {\bf 03} (2018) 109,
  [\href{http://arxiv.org/abs/1711.10391}{{\tt arXiv:1711.10391}}].

\bibitem{Peskin:1990zt}
M.~E. Peskin and T.~Takeuchi, {\it {A New constraint on a strongly interacting
  Higgs sector}},  {\em Phys. Rev. Lett.} {\bf 65} (1990) 964--967.

\bibitem{Barbieri:2004qk}
R.~Barbieri, A.~Pomarol, R.~Rattazzi, and A.~Strumia, {\it {Electroweak
  symmetry breaking after LEP-1 and LEP-2}},  {\em Nucl. Phys. B} {\bf 703}
  (2004) 127--146, [\href{http://arxiv.org/abs/hep-ph/0405040}{{\tt
  hep-ph/0405040}}].

\bibitem{Jenkins:2013wua}
E.~E. Jenkins, A.~V. Manohar, and M.~Trott, {\it {Renormalization Group
  Evolution of the Standard Model Dimension Six Operators II: Yukawa
  Dependence}},  {\em JHEP} {\bf 01} (2014) 035,
  [\href{http://arxiv.org/abs/1310.4838}{{\tt arXiv:1310.4838}}].

\bibitem{Fuentes-Martin:2022jrf}
J.~Fuentes-Mart\'\i{}n, M.~K\"onig, J.~Pag\`es, A.~E. Thomsen, and F.~Wilsch,
  {\it {A proof of concept for matchete: an automated tool for matching
  effective theories}},  {\em Eur. Phys. J. C} {\bf 83} (2023), no.~7 662,
  [\href{http://arxiv.org/abs/2212.04510}{{\tt arXiv:2212.04510}}].

\bibitem{Kraml:2025fpv}
S.~Kraml, A.~Lessa, S.~Prakash, and F.~Wilsch, {\it {SUSY meets SMEFT: Complete
  one-loop matching of the general MSSM}},
  \href{http://arxiv.org/abs/2506.05201}{{\tt arXiv:2506.05201}}.

\bibitem{Bellafronte:2023amz}
L.~Bellafronte, S.~Dawson, and P.~P. Giardino, {\it {The importance of flavor
  in SMEFT Electroweak Precision Fits}},  {\em JHEP} {\bf 05} (2023) 208,
  [\href{http://arxiv.org/abs/2304.00029}{{\tt arXiv:2304.00029}}].

\bibitem{Biekotter:2025nln}
A.~Biek{\"o}tter and B.~D. Pecjak, {\it {Analytic results for electroweak
  precision observables at NLO in SMEFT}},
  \href{http://arxiv.org/abs/2503.07724}{{\tt arXiv:2503.07724}}.

\bibitem{Draper:2011aa}
P.~Draper, P.~Meade, M.~Reece, and D.~Shih, {\it {Implications of a 125 GeV
  Higgs for the MSSM and Low-Scale SUSY Breaking}},  {\em Phys. Rev. D} {\bf
  85} (2012) 095007, [\href{http://arxiv.org/abs/1112.3068}{{\tt
  arXiv:1112.3068}}].

\bibitem{Perelstein:2007nx}
M.~Perelstein and C.~Spethmann, {\it {A Collider signature of the
  supersymmetric golden region}},  {\em JHEP} {\bf 04} (2007) 070,
  [\href{http://arxiv.org/abs/hep-ph/0702038}{{\tt hep-ph/0702038}}].

\bibitem{CMS:2021eha}
{\bf CMS} Collaboration, A.~Tumasyan et~al., {\it {Combined searches for the
  production of supersymmetric top quark partners in proton{\textendash}proton
  collisions at $\sqrt{s} = 13\,\text {Te}\text {V} $}},  {\em Eur. Phys. J. C}
  {\bf 81} (2021), no.~11 970, [\href{http://arxiv.org/abs/2107.10892}{{\tt
  arXiv:2107.10892}}].

\bibitem{ATLAS:2024itc}
{\bf ATLAS} Collaboration, G.~Aad et~al., {\it {ATLAS searches for additional
  scalars and exotic Higgs boson decays with the LHC Run~2 dataset}},  {\em
  Phys. Rept.} {\bf 1116} (2025) 184--260,
  [\href{http://arxiv.org/abs/2405.04914}{{\tt arXiv:2405.04914}}].

\bibitem{Baer:2025zqt}
H.~Baer, V.~Barger, J.~Bolich, J.~Dutta, D.~Martinez, S.~Salam, D.~Sengupta,
  and K.~Zhang, {\it {Prospects for supersymmetry at high luminosity LHC}},
  \href{http://arxiv.org/abs/2502.10879}{{\tt arXiv:2502.10879}}.

\bibitem{Haber:1996fp}
H.~E. Haber, R.~Hempfling, and A.~H. Hoang, {\it {Approximating the radiatively
  corrected Higgs mass in the minimal supersymmetric model}},  {\em Z. Phys. C}
  {\bf 75} (1997) 539--554, [\href{http://arxiv.org/abs/hep-ph/9609331}{{\tt
  hep-ph/9609331}}].

\bibitem{PardoVega:2015eno}
J.~Pardo~Vega and G.~Villadoro, {\it {SusyHD: Higgs mass Determination in
  Supersymmetry}},  {\em JHEP} {\bf 07} (2015) 159,
  [\href{http://arxiv.org/abs/1504.05200}{{\tt arXiv:1504.05200}}].

\bibitem{Arvanitaki:2011ck}
A.~Arvanitaki and G.~Villadoro, {\it {A Non Standard Model Higgs at the LHC as
  a Sign of Naturalness}},  {\em JHEP} {\bf 02} (2012) 144,
  [\href{http://arxiv.org/abs/1112.4835}{{\tt arXiv:1112.4835}}].

\bibitem{deBoer:1996hd}
W.~de~Boer, A.~Dabelstein, W.~Hollik, W.~Mosle, and U.~Schwickerath, {\it
  {Updated global fits of the SM and MSSM to electroweak precision data}},
  \href{http://arxiv.org/abs/hep-ph/9609209}{{\tt hep-ph/9609209}}.

\bibitem{Antusch:2015nwi}
S.~Antusch and C.~Sluka, {\it {Predicting the Sparticle Spectrum from GUTs via
  SUSY Threshold Corrections with SusyTC}},  {\em JHEP} {\bf 07} (2016) 108,
  [\href{http://arxiv.org/abs/1512.06727}{{\tt arXiv:1512.06727}}].

\bibitem{Marandella:2005wc}
G.~Marandella, C.~Schappacher, and A.~Strumia, {\it {Supersymmetry and
  precision data after LEP2}},  {\em Nucl. Phys. B} {\bf 715} (2005) 173--189,
  [\href{http://arxiv.org/abs/hep-ph/0502095}{{\tt hep-ph/0502095}}].

\bibitem{ATLAS:2024qxh}
{\bf ATLAS} Collaboration, G.~Aad et~al., {\it {Statistical Combination of
  ATLAS Run 2 Searches for Charginos and Neutralinos at the LHC}},  {\em Phys.
  Rev. Lett.} {\bf 133} (2024), no.~3 031802,
  [\href{http://arxiv.org/abs/2402.08347}{{\tt arXiv:2402.08347}}].

\bibitem{EWPPGTH}
\url{https://indico.cern.ch/event/1551879/}.
\newblock PPG EW WG meeting on TH uncertainties in EWPOs and Higgs for future
  $e^+ e^-$ colliders.

\bibitem{Selvaggi:2025kmd}
{\bf FCC} Collaboration, {\it {Prospects in electroweak, Higgs and Top physics
  at FCC}}, .

\end{thebibliography}\endgroup

\end{document}